\documentclass[aps,prl,floatfix,twocolumn,superscriptaddress]{revtex4}
\usepackage{amsmath,amssymb}
\usepackage{graphicx}
\usepackage{epsfig}
\usepackage{psfrag}
\usepackage[usenames]{color}
\usepackage{ulem}

\begin{document}









\title{Shedding light on topological superconductors}

\author{K.~H.~A.~Villegas}
\affiliation{Center for Theoretical Physics of Complex Systems, Institute for Basic Science, Daejeon, South Korea}

\author{V.~M.~Kovalev}
\affiliation{A.V. Rzhanov Institute of Semiconductor Physics, Siberian Branch of Russian Academy of Sciences, Novosibirsk 630090, Russia}
\affiliation{Department of Applied and Theoretical Physics, Novosibirsk State Technical University, Novosibirsk 630073, Russia}

\author{F.~V.~Kusmartsev}
\affiliation{Department of Physics, Loughborough University, Loughborough LE11 3TU, United Kingdom}

\author{I.~G.~Savenko}
\affiliation{Center for Theoretical Physics of Complex Systems, Institute for Basic Science, Daejeon, South Korea}


\date{\today}

\begin{abstract}
We propose an optical approach to monitor superconductors in conjunction with a normal metal layer. 
Effectively such a hybrid system represents a resonator, where electrons are strongly coupled with light.
We show that the interaction of light with the superconductor turns out strongly boosted in the presence of the neighboring metal 
and as a result, the electromagnetic power absorption of the system is dramatically enhanced.
It manifests itself in a giant Fanolike resonance which can uniquely characterize the elementary excitations of the system. 
Our approach is especially promising for topological superconductors, where the Majorana fermions can be revealed and controlled by light.
\end{abstract}


\maketitle


Superconductivity is conventionally considered to be a material property difficult to characterize with light, due to weak light--matter interaction in superconducting condensates. To test whether materials are superconductors, electric (resistivity-based) and magnetic (Meissner effect-based) techniques are routinely used. However, optics is required if we want to monitor hybrids of such fascinating classes of materials as topological insulators (TI)~\cite{kane2005z,kane2005quantum,bernevig2006quantum,fu2007topological,kusmartsev1985semimetallic} 
and Weyl semimetals (WM)~\cite{son2013chiral,lv2015experimental,soluyanov2015type,yang2015weyl,xu2015discovery,chiu2016classification,sun2015prediction,zhang2016signatures}. In this framework, superconductors should be considered as candidates to reveal new topological properties, thus fostering a revisit of existing experimental techniques.

Indeed, there is a strong current demand to develop alternative methods of diagnostics. Topological superconductors behaving metallic on the surface and superconducting in the bulk naturally combine the properties of both metal and superconductor, which is the key problem for their discovery. On one hand, electrical conductivity measurements used to study conventional superconductors have proved to be challenging due to a mutual influence of the free electrons associated with the metallic surface and the Cooper pairs of the superconducting bulk. On the other hand, diamagnetic Meissner magnetization measurements require a minimal volume, which is an issue for surface topological superconductivity.

In this Letter, we propose an optical approach that is seemingly ideal for a general metal--superconducting hybrid setup. 
As an important ingredient, it requires adding to the system an extra metallic layer with plasmonic gapless excitations. Their hybridization with flat bands and superconducting excitations leads to a giant light absorption due to a high density of states. As a result, a film of topological material deposited on the thin metallic layer will have hybrid elementary excitations. 
The significant result that we demonstrate below is that these 
excitations are highly optically active; 
they display strong resonances in optical absorption measurements, thereby characterizing the material under study.

\begin{figure*}[!t]
\centering
\includegraphics[width=18cm]{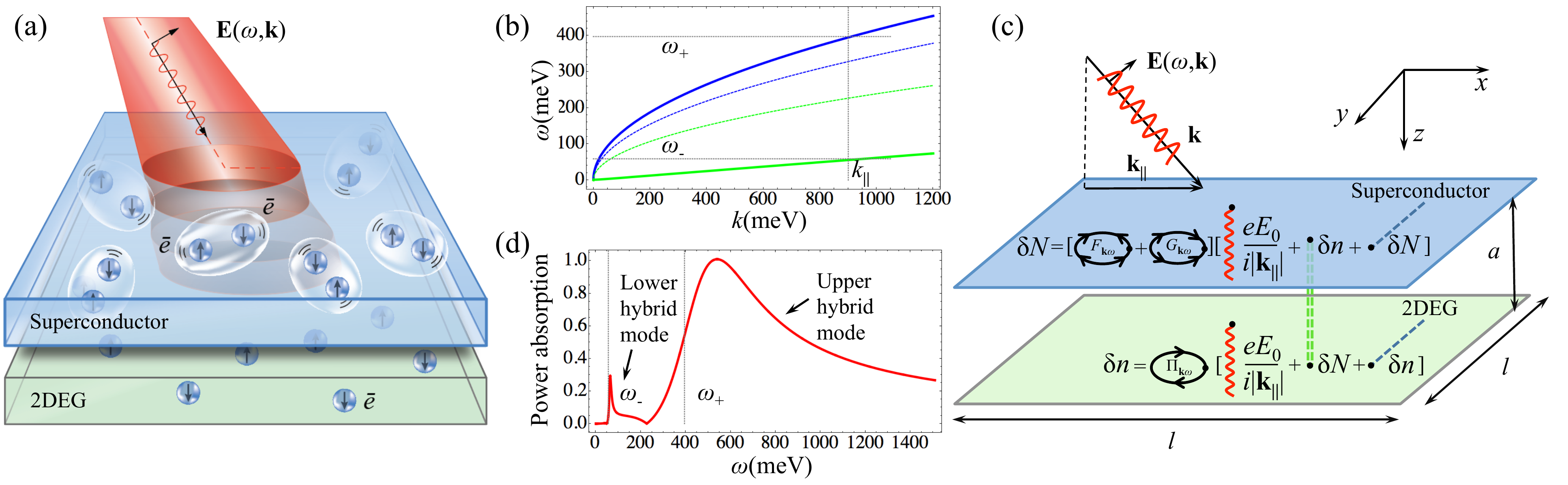}
\caption{\label{Fig1} 
System schematic. 
(a) Hybrid normal metal--superconductor structure exposed to an electromagnetic field of incident light.
(b) Dispersions of hybrid eigenmodes of the system: $\omega$ as a function of $k$ for $m_N=m_S=1$ (green and blue solid curves). The dashed curves of the corresponding colors show the individual modes of each layer when the interlayer interaction is switched off. 
(c) Schematic of in-layer, inter-layer and light-matter interaction in the system manifesting itself in fluctuations of electron and Cooper pair densities, $\delta n$ and $\delta N$, and polarization operators $F_{\mathbf{k}\omega}$, $G_{\mathbf{k}\omega}$, and $\Pi_{\mathbf{k}\omega}$.
(d) Spectrum of electromagnetic power absorption demonstrating the Fano resonance profile.}
\end{figure*}

Various hybrid normal metal--superconductor or semiconductor--superconductor systems, in which a two-dimensional electron gas (2DEG) is in contact with Cooper pairs, have been broadly considered in literature. Examples of the widespread implementation of such hybrid systems include a Josephson junction or a Josephson tunnel junction~\cite{VanDuzer} aimed at cooling (as a heat sink), observation of Majorana fermions~\cite{Mourik,Takei2013} and zero modes~\cite{Takei, Suominen}, and an enhancement of the degree of photon pair entanglement~\cite{Khoshnegar}. 



Recently, it was shown that hybrid systems also allow new mechanisms of superconductivity itself using interaction with excitons~\cite{Skopelitis} or exciton polaritons~\cite{Shelykh, Imamoglu} in semiconductor structures, where the latter can serve as an auxiliary to increase $T_c$.
From an application-oriented perspective, semiconductor-based hybrid structures can be employed in such devices as tunnel diodes~\cite{Hayat} and optoelectronic circuits for high-bandwidth information processing~\cite{Shabani, Shainline}. Furthermore, the most recent advances in molecular beam epitaxial heterostructure growth techniques suggest a route to create high-quality hybrid structures~\cite{Yan}.

We show here that a metallic layer located in the vicinity of a superconductor can dramatically enhance light-superconductor coupling, such that the superconducting properties can ultimately be well characterized by the absorption spectrum of the hybrid system. Specifically, we demonstrate that the system reveals a giant hybrid Fano resonance~\cite{FanoPaper, RefOurPRB}, which arises in both normal and superconducting hybrid subsystems due to their mutual influence.


The shape and positions of the peaks (and the dip) of the Fano resonance may uniquely characterize both the superconducting and metallic subsystems, especially the value of the superconducting gap and therefore the order parameter, its symmetry and critical temperature. Thus, our findings open a prospective  method, being optical and noninvasive, for the characterization and testing of materials for superconductivity.


We consider a system with two parallel layers of a normal metal and a superconductor, as illustrated in Fig.~\ref{Fig1}a. 
The electrons in the normal metal interact via Coulomb interaction, which has the Fourier image given by $v_k=2\pi e^2/k$, where $\mathbf{k}$ is in-plane momentum (lying in the $xy$ plane).
The electrons between the two layers are also Coulomb-coupled, and the Fourier image of the interlayer interaction reads $u_k=2\pi e^2\exp(-ak)/k$, where $a$ is the separation between the layers. The electromagnetic wave is polarized along the $x$-axis, 
$\mathbf{E}(\mathbf{r},t)=\mathbf{\hat{x}}E_0e^{-i(k_{\perp}z+\mathbf{k}_\parallel\cdot\mathbf{r}+\omega t)}$
where $\mathbf{k}_\parallel$, $\omega$, and $\mathbf{r}$ are the in-plane wave vector of the field, frequency, and coordinate, respectively. 

\begin{figure*}[t!]
\includegraphics[width=16cm]{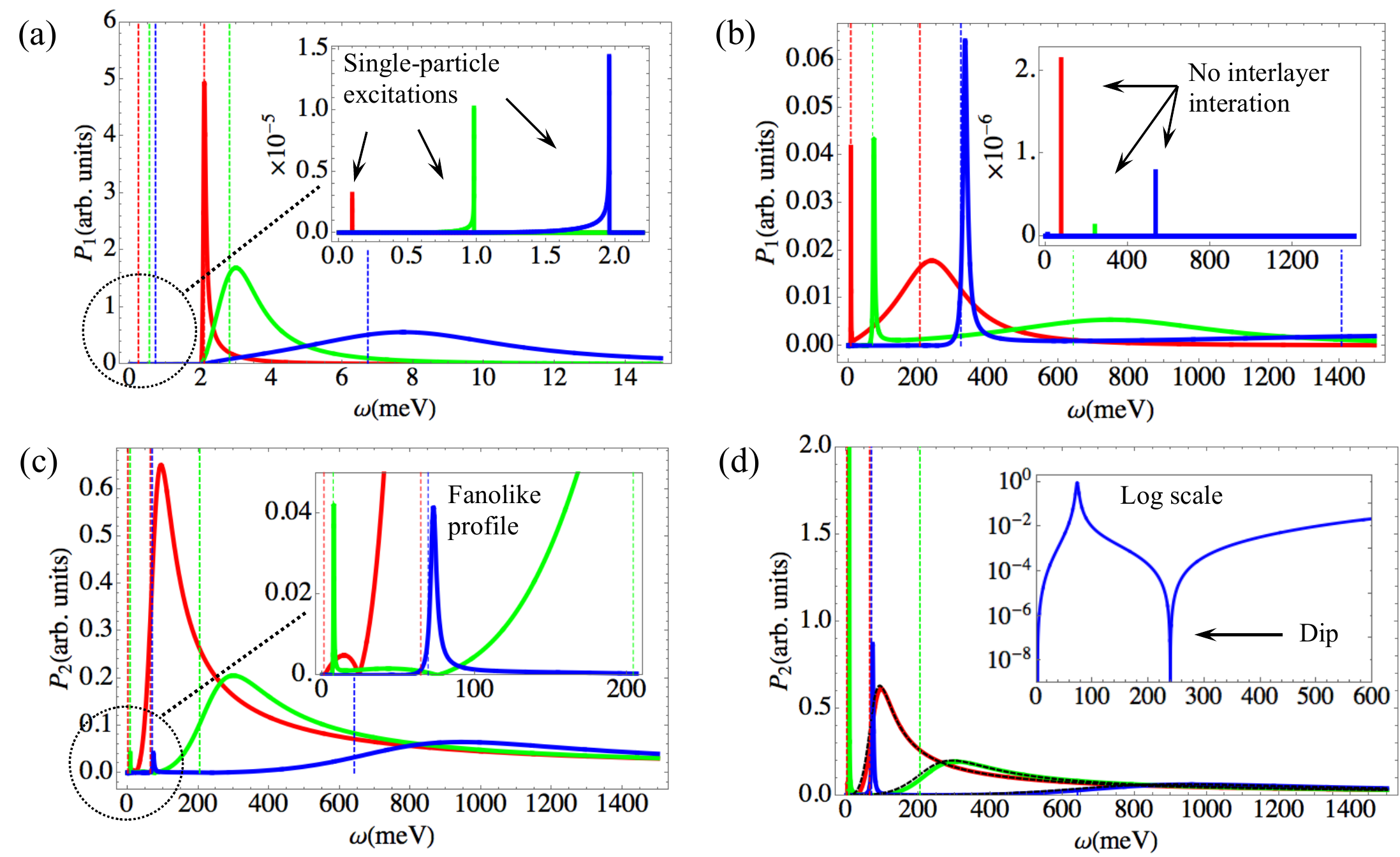}
\caption{\label{Fig2} 
Spectra. Power absorption monitored in 2DEG ((a) -- (b)) and superconductor ((c)--(d)) as a function of $\omega$ for $\Delta=1.0$ meV. 
Vertical dashed lines stand for the corresponding locations of the hybrid modes (within Eq.~(\ref{Eq1})).
(a) $k=1.0\times 10^{-3}$ (red curve), $1.0\times 10^{-2}$ (green curve), and $1.0\times 10^{-1}$ meV (blue curve).  
Inset shows the range $0\leq\omega<2\Delta$, contributions to $P_1(\omega)$ due to single-particle excitations. To render these contributions visible, larger $k$'s were used: $k=5.0\times 10^{1}$ (red curve), $5.0\times 10^{2}$ (green curve), and $1.0\times 10^{3}$ (blue curve). 
(b) $k=1.0\times 10^{1}$ (red curve), $1.0\times 10^{2}$ (green curve), and $1.0\times 10^{3}$ meV (blue curve). 
Inset shows the case with no interlayer interaction.
(c) Both layers are exposed to the EMF. Inset: zoom-in for small $\omega$'s showing peaks caused by the lower hybrid modes.
(d) No external field on the normal layer. Inset shows the log plot of the corresponding blue curve, manifesting the two peaks and the dip of the Fano resonance.
In (c) and (d),
$k=1.0\times 10^{1}$ (red curves), $1.0\times 10^{2}$ (green curves), $1.0\times 10^{3}$ meV (blue curves). Dashed black curves in (d) show the case when interlayer coupling is turned off.}
\end{figure*}
%
%
%


%
%
%
%
%


Using the polarization functions of the 2DEG and superconductor (see Supplemetal Material~\cite{SM} and~\cite{Arseev,Fetter}), we solve the eigenvalue problem and find two branches of dispersion of the hybrid modes (assuming a topologically trivial case of constant $\Delta$):
\begin{eqnarray}\label{Eq1}
\omega^2_{\pm}(k)&=&2\Delta^2+\frac{e^2k}{2}\left(\frac{p^2_{NF}}{\pi m_N}+\frac{2p_{SF}^2}{m_S}\right)\pm
\\
\nonumber
&&~~~~\pm\frac{1}{2}
\sqrt{\xi_{-}\xi_{+}
-4\beta_k^2},~~~\textrm{where}\\
\nonumber
&&\xi_{\pm}=\left[\left(2\Delta\pm ep_{NF}\sqrt{\frac{k}{\pi m_N}}\right)^2+\frac{2e^2}{m_S}p_{SF}^2k\right],
\\
\nonumber
&&\beta^2_k=\frac{2e^4p_{NF}^2p_{SF}^2}{\pi m_Nm_S}k^2(1-e^{-2ka}).
\end{eqnarray}
Here, $2\Delta$ is the superconducting gap, $e>0$ is elementary charge, $p_{SF}$ and $p_{NF}$ are the Fermi momenta in the superconductor and normal layer, respectively, and $m_S$ and $m_N$ are effective electron masses.
Figure~\ref{Fig1}b shows the hybrid modes for various $m_N=m_S$ (with bare modes presented for comparison). 
%


Further, assuming a linear response of the system, the Fourier component of the current in the 2DEG layer can be written as $j_{k_{\parallel},\omega}$. Both the wave vector and frequency of current density have specific values fixed by the external electromagnetic field;
the following formula can therefore be used to compute the time-averaged power absorbed by the hybrid system as a function of frequency $\omega$:
\begin{eqnarray}
\nonumber
{\cal P}(\omega)&=&\frac{1}{2}\left\langle\mathcal{R}e\int d^2r\mathbf{J}(\mathbf{r},t)\cdot \mathbf{E}^*(\mathbf{r},t)\right\rangle,
\end{eqnarray}
where the integration is over the plane of the normal metal sample and $\langle\cdot\cdot\cdot\rangle$ denotes time-averaging.
Normalizing ${\cal P}(\omega)$ by $\int d^2r=l^2$ and utilizing the continuity equation $kj_{k,\omega}=-e\omega\delta n_{k,\omega}$, where $\delta n_{k,\omega}$ describes fluctuations of the electron density in the 2DEG (see the explicit formula in Supplemental Material~\cite{SM}, section I.A), we obtain the specific power absorption coefficient (hereafter simply referred to as \textit{power absorption}):
\begin{eqnarray}
P_1(\omega)&=&\frac{1}{2}\cdot\frac{e\omega}{k}\left|\mathcal{R}e(\delta n_{k,\omega})\right|E_0.
\end{eqnarray}
This formula accounts for the electron-electron and electron-Cooper pair interaction as well as the coupling of both the 2DEG and superconductor to light (see the schematic description of corresponding processes in Fig.~\ref{Fig1}c).

%


%
%
%
%
%
%


The resulting power absorption by the hybrid system is presented in Fig.~\ref{Fig1}d, where due to the interplay of different interaction mechanisms, we observe a Fano resonance. Let us consider the spectrum in detail.
Figure~\ref{Fig2} shows power absorption as a function of $\omega$ for different wave vectors $k$ when both the 2DEG and superconductor are exposed to the EMF (the situation where the superconductor is transparent is considered in Supplemental Material~\cite{SM}). 
In Fig.~\ref{Fig2}a, the lower hybrid modes are below $2\Delta$ and their contribution to the power absorption is suppressed, as can be seen by the lack of visible peaks in the vicinity of the three left-most dashed lines. The inset shows the contribution of the single-particle excitations. As can be seen, this contribution is negligible compared to the contribution of the hybrid modes (three peaks of the main plot). The locations of the peaks at higher frequencies nearly coincide with the corresponding dashed lines, showing that these peaks are primarily due to the upper hybrid modes. As $k$ increases, we observe a broadening (from red to blue curves).

Figure~\ref{Fig2}b shows the power absorption for larger values of $k$. The lower hybrid modes now have significant contributions, as can be seen by the existence of three sharp peaks, since they are now located above the gap. For comparison, the inset shows the power absorption when the interlayer coupling is switched off. It shows a disappearance of the contribution from the upper hybrid modes primarily due to the superconductor. Comparison of the lower-mode contributions shows that the presence of the superconductor enhances the power absorption of the hybrid system by four orders of magnitude. 
Otherwise, the presence of the external EMF and the 2DEG does not dramatically influence the superconductor itself, and thus may serve as an auxiliary to monitor its behavior.


We can follow a similar procedure to calculate Cooper pair current in the superconductor. The power absorption then reads (see Supplemental Material~\cite{SM}, section I.B):
\begin{eqnarray}
P_2(\omega)&=&\frac{1}{2}\cdot\frac{2e\omega}{k}|\mathcal{R}e(\delta N_{k\omega})|E_0,
\end{eqnarray}
%
%
where $\delta N_{k\omega}$ are Cooper pair density fluctuations in the superconducting layer.

Figure~\ref{Fig2}c shows the power absorption spectrum for different $k$'s when the EMF is exposed to both the 2DEG and superconductor. We observe a celebrated Fano resonance structure of the spectrum (see Inset).
Figure~\ref{Fig2}d shows the power absorption when the external EMF at the normal layer is turned off. We immediately note that the contribution of the upper hybrid modes to the absorption spectrum is not significantly affected by switching off the field at the normal layer, as can be seen by comparing Figs.~\ref{Fig2}c and~\ref{Fig2}d. This behavior is quite expected. Surprisingly though, the contribution of the lower hybrid modes shows significant increase compared to the situation when the external field in the normal layer is switched on. Thus, we realize that the electronic layer must be very sensitive to the behavior of the superconductor, whereas the latter does not pay much attention to either the 2DEG or the light field.
Indeed, switching off the interlayer interaction removes the contribution of the lower modes, as expected (dashed black curves in Fig.~\ref{Fig2}d). We see that the second peak, which is mostly determined by the superconductor, remains nearly the same. Inset shows that in log scale, we observe both the two peaks of the Fanolike resonance and the dip.
\begin{figure}[t!]
\includegraphics[width=8.5cm]{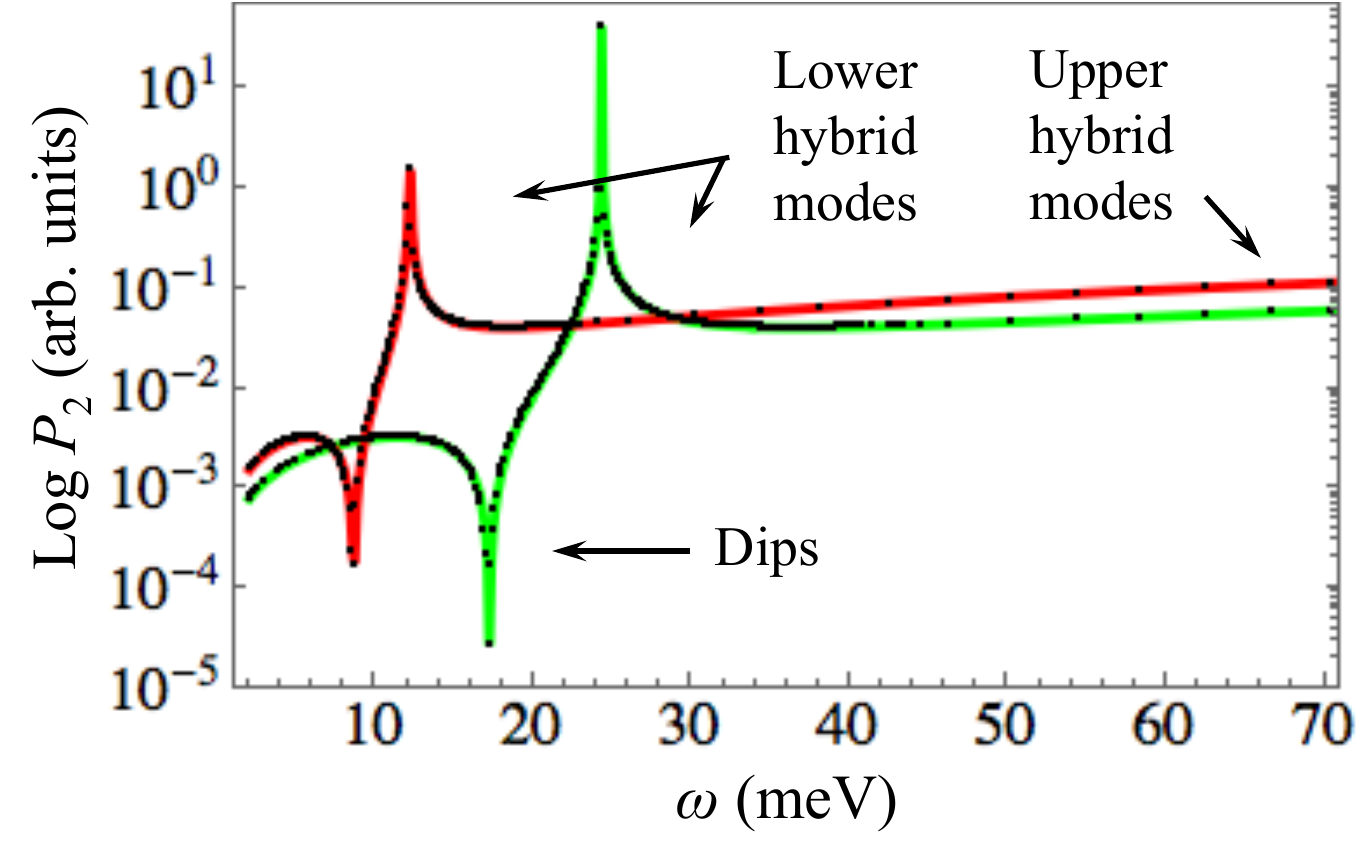}
\caption{\label{Fig3} Spectrum in topologically non-trivial case. Power absorption (in log scale) monitored in superconductor as a function of $\omega$ at $k=0.5\times 10^{-1}$ meV (red) and $k=1.0\times 10^{-1}$ meV (green). (Black dots show the results of numerical calculation.)}
\end{figure}

Now let us consider the nontrivial topology (p-wave pairing). Then the superconducting gap reads $\Delta_\mathbf{k}=f(k)(k_x-ik_y)=f(k)ke^{-i\phi}$, where $\phi$ is the polar angle measured with respect to the $k_x$ axis in k-space and the function $f(k)$ depends on the concrete p-wave superconductor;  it should be finite for all $k\in [0,\infty)$ and vanish as $k\rightarrow\infty$ (see all the calculations in Supplemental Material~\cite{SM}, Section IIC, D). Dependence $\Delta_\mathbf{k}$ makes calculations more tricky but leads to qualitatively similar results. Fig.~\ref{Fig3} shows the resulting $P_2$ (compare with Fig.~\ref{Fig2}c, green curve). It should be noted, that here the Fanolike dip is shifted to the frequency range below the first peak.


\textit{Conclusions.---}
We have studied the linear response of a hybrid two-dimensional electron gas--superconductor system to an external electromagnetic field of light.
Such systems have hybrid excitations that originate from gapless plasmons of the two-dimensional electron gas of the metallic layer and gapfull Bogolubov excitations of the bulk superconductor. 
We calculated these hybrid eigenmodes of the system and investigated the electromagnetic power absorption spectra. We found that these excitations exhibit a very strong coupling with the electromagnetic radiation, and showed that they display giant Fano resonances 
 associated with a large light absorption.  
Such results therefore indicate a way to monitor the behavior of a superconductor exposed to light by measuring the  spectrum of photoabsorption of the two-dimensional electron gas.

Herewith, we suggest a way to prepare samples of various hybrid-structured materials to test them for being superconductors via optics. That is, the proposed effect of the giant hybrid Fano resonance can be observed by measuring the optical response of two-layer metal-superconductor systems composed of, for example, superconducting niobium film deposited on a thin metallic layer of copper, gold, silver, or tin. The proposed effect can be used to design various sensors and diagnostics of superconducting magnets with light.

Moreover, the recent discovery of high-temperature light-induced superconductivity in K$_2$C$_{60}$~\cite{Nature530} has stimulated an  activity in the scientific community to test materials with light. Thus, our finding of enhanced light coupling in metal--superconducting hybrids alongside the possibility of testing is expected to open a new direction in this activity, since it creates many opportunities for discoveries of condensed states induced by light.


We thank Sergej Flach for useful discussions, Joel Rasmussen (RECON) for a critical reading of our manuscript, and Ekaterina Savenko for help with the figures.
V.~M.~K. has been supported by the Russian Foundation for Basic Reaserch (Project No.~16-02-00565).
K.~H.~V. and I.~G.~S. acknowledge the support of the Institute for Basic Science in Korea (Project No.~IBS-R024-D1).





%
%




%
%
%
%
%
%
%
%

%
%

\end{document}